\documentclass[useAMS,usenatbib,usegraphicx]{mn2e}

\title[Black holes in massive star clusters]{The effect of stellar-mass black holes on the 
structural evolution of massive star clusters}
\author[A.~D.~Mackey et al.]{A.~D.~Mackey$^{1}$, M.~I.~Wilkinson$^{2}$, 
M.~B.~Davies$^{3}$, and G.~F.~Gilmore$^{4}$\\
$^{1}$Institute for Astronomy, University of Edinburgh, Royal Observatory, Blackford 
Hill, Edinburgh, EH9 3HJ, UK\\
$^{2}$Department of Physics \& Astronomy, University of Leicester, University Road,
Leicester, LE1 7RH, UK\\
$^{3}$Lund Observatory, Box 43, SE-221 00 Lund, Sweden\\
$^{4}$Institute of Astronomy, University of Cambridge, Madingley Road, Cambridge, 
CB3 0HA, UK}
\begin{document}

\date{Accepted 2007 April 19. Received 2007 April 18; in original form 2007 March 26.}

\pagerange{\pageref{firstpage}--\pageref{lastpage}} \pubyear{2007}

\maketitle

\label{firstpage}

\begin{abstract}
We present the results of realistic $N$-body modelling of massive star clusters
in the Magellanic Clouds, aimed at investigating a dynamical origin for the radius-age
trend observed in these systems. We find that stellar-mass black holes, formed in the
supernova explosions of the most massive cluster stars, can constitute a dynamically 
important population. If a significant population is retained (here we assume complete 
retention), these objects rapidly form a dense core where interactions are 
common, resulting in the scattering of black holes into the cluster halo, and 
the ejection of black holes from the cluster. These two processes heat 
the stellar component, resulting in prolonged core expansion of a magnitude matching the 
observations. Significant core evolution is also observed in Magellanic Cloud clusters at 
early times. We find that this does not result from the action of black holes, 
but can be reproduced by the effects of mass-loss due to rapid stellar evolution in a 
primordially mass segregated cluster. 
\end{abstract}

\begin{keywords}
stellar dynamics -- globular clusters: general -- methods: $N$-body simulations
-- Magellanic Clouds.\vspace{-4mm}
\end{keywords}

\section{Introduction}
Globular clusters are central to a wide variety of astrophysical research, 
ranging from star formation, stellar and binary star evolution, and stellar dynamics, 
through to galaxy formation and evolution, and cosmology. These objects therefore 
constitute an integral part of our understanding of the Universe, and it is clearly 
vital that their internal evolutionary processes are well understood.
The Galactic globular clusters, while close, are exclusively ancient objects ($\tau \ga 10^{10}$ 
yr). We can therefore accurately determine the end-points of their evolution, 
but must infer the complete long-term development which brought them to these observed 
states. To directly observe cluster evolution, we must switch our attention to the 
Magellanic Clouds (LMC/SMC), which both possess extensive systems of 
star clusters with masses comparable to the Galactic globulars, but crucially {\it 
of all ages:} $10^6 \la \tau \la 10^{10}$ yr. These systems are of fundamental 
importance because they are the nearest places we can observe snapshots of all phases 
of cluster development.

\citet*{elson:89}, discovered a striking relationship between core radius ($r_c$) and age for 
LMC clusters -- namely that the observed spread in $r_c$ increases dramatically with increasing 
age. Here, $r_c$ is the observational core radius, defined as the projected radius at 
which the surface brightness has decreased to half its central value.
Recently, \citet{mackey:03a,mackey:03b} used Hubble Space Telescope (HST) WFPC2 imaging 
of $63$ massive Magellanic Cloud clusters to more clearly demonstrate the radius-age trend
in the LMC and show, for the first time, that a radius-age trend also exists 
in the SMC. An additional $46$ objects have since been observed with HST/ACS 
(Program \#9891) to improve sampling of the radius-age plane. Structural measurements for all 
$107$ clusters may be seen in Fig \ref{f:pair1}.

The observed radius-age relationship provides strong evidence that our understanding
of globular cluster evolution is incomplete, since standard quasi-equilibrium models 
do not predict large-scale core expansion spanning a full cluster life-time 
\citep[see e.g.,][]{meylan:97}. Discerning the origin of the radius-age trend is 
therefore of considerable importance. A number of groups have investigated possible 
explanations -- these include a size-of-sample bias \citep{hunter:03}, heating due to 
binary stars or tidal shocks \citep{wilkinson:03}, and the formation of cores in 
primordially cusped clusters due to the sinking of massive stellar remnants \citep{merritt:04}. 
However, a model which fully accounts for the observed trend has yet to be elucidated.

The radius-age trend is indistinguishable in the LMC and SMC, and the oldest LMC/SMC 
clusters have $r_c$ distributions consistent with those of globular clusters in our
Galaxy and in the Fornax and Sagittarius dSph galaxies \citep{mackey:04}.
Since these galaxies have very different tidal fields and possible external torques, this
suggests that the radius-age relation is driven by internal cluster processes, 
with any external or tidal effects second order \citep[see also][]{wilkinson:03}.
In this Letter, we report on the results of direct, realistic $N$-body simulations 
designed to investigate an internal dynamical origin for the radius-age trend. 
We follow the structural evolution of 
model clusters with varying degrees of primordial mass segregation (MSeg), possessing 
populations of stellar-mass black holes (BHs) formed in the supernova explosions of 
the most massive cluster stars.
We demonstrate that a cluster which retains its BHs undergoes dramatic core 
expansion for most of its lifetime, in contrast to a cluster with no BHs, which proceeds 
towards core collapse. We also show that primordial MSeg has an important effect on the 
early evolution of a cluster, when mass-loss due to stellar evolution is severe.

\vspace{-5mm}
\section{Numerical setup}
\label{s:nbodycode}
Direct $N$-body modelling is a powerful tool for studying star cluster evolution
because it incorporates all relevant physics with a minimum of simplifying
assumptions. We have used the {\sc nbody4} code in combination with a 32-chip GRAPE-6 
special-purpose computer \citep{makino:03} to run simulations of Magellanic Cloud 
clusters. Full details of {\sc nbody4} are provided by \citet{aarseth:03}.
It uses a fourth-order Hermite scheme and evaluation of the 
force and its first time derivative by the GRAPE-6 to integrate the equations of motion. 
Close encounters between stars, including stable binary systems, are 
treated with two-body or chain regularization algorithms. Also incorporated are routines for 
modelling the stellar evolution of single and binary stars \citep{hurley:00,hurley:02}. 
These include a metallicity dependence, and a mass-loss prescription 
such that evolving stars lose gas through winds and supernova explosions. 

\begin{table*}
\begin{minipage}{164mm}
\caption{Details of $N$-body runs and initial conditions. Each cluster begins with 
$N_0$ stars with masses summing to $M_{{\rm tot}}$, and initial central density 
$\rho_0$. Initial cluster structure is ``observed'' to obtain $r_c$ and $\gamma$. 
Each model is evolved until $\tau_{{\rm max}}$.}
\centering
\begin{tabular}{@{}lccccccccccc}
\hline \hline
Name & \hspace{0mm} & $N_0$ & $\log M_{{\rm tot}}$ & $\log \rho_0$ & $r_c$ & $\gamma$ & \hspace{0mm} & Initial MSeg & BH Retention & \hspace{0mm} & $\tau_{{\rm max}}$ \\
 & & & (${\rm M}_\odot$) & (${\rm M}_\odot\,{\rm pc}^{-3}$) & (pc) & & & ($T_{{\rm MS}}$) & ($f_{{\rm BH}}$) & & (Myr) \\
\hline
Run 1 & & $100\,881$ & $4.746$ & $2.31$ & $1.90 \pm 0.09$ & $2.96 \pm 0.17$ & & None & $0.0$ & & $12\,000$ \\
Run 2 & & $100\,881$ & $4.746$ & $2.31$ & $1.90 \pm 0.09$ & $2.96 \pm 0.17$ & & None & $1.0$ & & $10\,668$ \\
\hline
Run 3 & & $95\,315$ & $4.728$ & $4.58$ & $0.25 \pm 0.04$ & $2.33 \pm 0.10$ & & $450$ Myr & $0.0$ & & $11\,274$ \\
Run 4 & & $95\,315$ & $4.728$ & $4.58$ & $0.25 \pm 0.04$ & $2.33 \pm 0.10$ & & $450$ Myr & $1.0$ & & $10\,000$ \\
\hline
\label{t:runs}
\end{tabular}
\vspace{-5mm}
\end{minipage}
\end{table*}

We generate models with initial properties as close as possible to those observed 
for young Magellanic Cloud clusters. These objects possess radial surface brightness (SB)
profiles best described by \citet*[][ hereafter EFF]{elson:87} models: 
$\mu(r) = \mu_0(1+r^2/a^2)^{-\gamma/2}$, where $\mu_0$ is the central SB, $a$ is the
scale length, and $\gamma$ the power-law fall-off at large $r$. Typically, their
core radii $r_c = a(2^{2/\gamma}-1)^{1/2} \sim 0.2\,$-$\,2.5$ pc, and $\gamma \sim 2.0\,$-$\,3.5$
\citep[e.g.,][]{mackey:03a}. Their central densities $\rho_0$\ (${\rm M}_{\odot}$pc$^{-3}$) 
lie in the range $1.5 \la \log \rho_0 \la 2.5$ (except for R136 which is much denser with 
$\log \rho_0 \sim 4.8$), while their total masses $M_{{\rm tot}}$\ (${\rm M}_{\odot}$)
lie in the range $4.0 \la \log M_{{\rm tot}} \la 5.6$ \citep{mclaughlin:05}.
We generate non-MSeg clusters by selecting stellar positions randomly from the density 
distribution of an EFF model with $\gamma=3$. Each star is assigned a velocity 
drawn from a Maxwellian distribution, where the velocity dispersion $\sigma$ is calculated 
using the Jeans equations assuming an isotropic velocity distribution. Expressions for 
$\sigma$ are given in a forthcoming paper (Mackey et al. 2007a, in prep.).
We select the initial mass function (IMF) of \citet{kroupa:01}, 
with a stellar mass range $0.1 - 100\,{\rm M}_\odot$. 
Choosing $N\sim10^5$ particles results in cluster masses of $\log M_{{\rm tot}} \sim 4.75$. 
We adopt $[$Fe$/$H$] = 0$, similar to young LMC clusters. However,
both Clouds exhibit strong age-metallicity relationships -- this may have important implications
for our results.

Resolved observations of very young Magellanic Cloud clusters invariably reveal some
degree of MSeg \citep[e.g.,][]{degrijs:02}. Detailed cluster formation models support 
such observations \citep[e.g.,][]{bonnell:06}; we would therefore like to
include the effects of MSeg in our modelling. We have developed a method to generate 
clusters with primordial MSeg in a ``self-consistent'' fashion; again, full details will be
provided by Mackey et al. (2007a, in prep.). Briefly, we take a non-MSeg 
cluster and use {\sc nbody4} to evolve it without stellar evolution, so that the 
cluster begins to dynamically relax. The degree of primordial MSeg is controlled via 
the duration of this ``pre-evolution'', $T_{{\rm MS}}$. The positions and 
velocities of the stars in the pre-evolved cluster are then used as the initial
conditions ($\tau=0$) for a full run with stellar evolution included.
Stars slowly escape during pre-evolution, so our MSeg models are marginally less massive 
than non-MSeg models. We generate MSeg clusters with $T_{{\rm MS}} = 450$ Myr. 
These models have structural properties (e.g., density profile, and radial mass-function
variation) consistent with those observed for very young Magellanic Cloud clusters. 

LMC clusters are observed at galactocentric radii spanning $\sim 0-14$ kpc; our models 
move on circular orbits of radius $6$ kpc about a point-mass LMC with 
$M_{\rm g} = 9\times 10^9 {\rm M}_\odot$. \citet{wilkinson:03} describe the implementation 
of an external tidal field within {\sc nbody4}. Adopting a point-mass LMC is an 
over-simplification; however \citet{wilkinson:03} showed
that a weak external field does not result in strong core evolution -- hence, here we are
only interested in internal processes. Clusters are assumed to initially just fill their 
tidal radii. The initial tidal radius of a model cluster therefore sets the ratio between 
the length units used by {\sc nbody4} \citep[see][]{aarseth:03}
and physical length units (pc). This scaling controls the physical density of the cluster and 
hence the physical time-scale on which internal dynamical processes occur. Our non-MSeg clusters 
have central density $\log \rho_0 = 2.31$ and core radius $r_c = 1.90$ pc, which matches 
typical young LMC and SMC objects. The primordially MSeg clusters have 
$\log \rho_0 = 4.58$ and $r_c = 0.25$ pc, which closely resembles the compact, massive
LMC cluster R136. Given this correspondence, we are confident in our selection
of an appropriate length-scale.

We have modified {\sc nbody4} to control the production of BHs in supernova
explosions. We can vary the minimum mass of a BH progenitor 
star, the masses of the BHs themselves, and the natal 
velocity kicks they receive. This is implemented in a simple
but serviceable manner. All stars initially above $20\,{\rm M}_\odot$ produce BHs,
with masses uniformly distributed in the range $8 \leq m_{{\rm BH}} \leq 12\,{\rm M}_\odot$. 
This range is consistent with dynamical masses obtained from observations of X-ray binaries
\citep[e.g.,][]{casares:06}. Each model cluster has the same random seed and so each begins with
an identical stellar population: our adopted IMF and total $N$ lead to the formation of
$198$ BHs in all clusters. In our models, natal BH kicks are either much larger than 
the cluster escape velocity $v_{{\rm esc}}$ (i.e., BH retention fraction 
$f_{{\rm BH}} = 0$) or zero ($f_{{\rm BH}} = 1$).

To obtain structural measurements consistent with those for real clusters, we
simulate observations of our $N$-body models. That is, we mimic the reduction 
procedures from which the HST $r_c$ measurements were derived.
In those observations, the bright (saturation) and faint (background-limited) stellar
detection levels are a weak function of cluster age, reflecting the requirement for 
longer exposure durations to image main sequence stars in older clusters. This is
not responsible for the radius-age trend, but must be accounted for 
in our analysis. Further, the WFPC2 and ACS fields-of-view limit radial profiles to 
a maximum extent of $\sim 70-100$ arcsec. To simulate these observations we first convert 
the luminosity and effective temperature of each $N$-body star to magnitude and colour
using the model atmospheres of \citet{kurucz:92} and \citet{bergeron:95}. Next, 
we impose appropriate bright and faint detection limits along with the field-of-view limits. 
We use the remaining stars to construct a SB profile, following \citet{mackey:03a}.
Stellar positions are projected onto a plane, and the SB calculated in circular annuli
about the cluster centre. A varying annulus width is used to evenly 
sample both the cluster core and halo. Finally, we fit an EFF model to the resulting 
profile to derive $r_c$ and $\gamma$. To reduce noise we average the results for
three orthogonal planar projections. 

\vspace{-5mm}
\section{Simulations and Results}
\label{s:results}
The parameter space of interest is spanned by non-MSeg clusters and those 
with significant primordial MSeg. For each of these, we consider evolution with BH retention
fractions $f_{{\rm BH}} = 0$ (large natal kicks) and $1$ (no natal kicks). These four runs define 
the extremities of the parameter space, and hence are expected to cover the limits of cluster 
behaviour. Their properties are listed in Table \ref{t:runs}. No special significance should be
attached to $\tau_{{\rm max}}$ -- these simply represent the most convenient termination
points for each simulation after $\tau = 10$ Gyr had been reached.

\vspace{-4mm}
\subsection{$N$-body pair 1: No mass segregation}
\label{ss:pair1}

\begin{figure}
\centering
\vspace{-2mm}
\includegraphics[width=8cm]{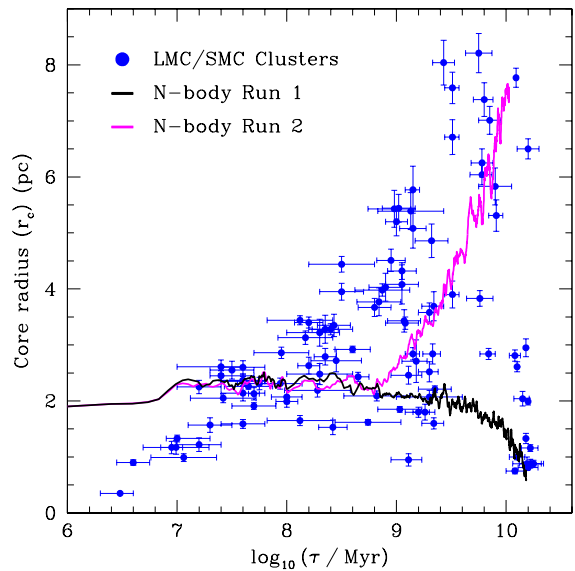}
\caption{Core radius evolution of Runs 1 and 2, which are initially identical,
with no primordial MSeg. They have $f_{{\rm BH}} = 0$ and $1$, respectively. 
Run 1 evolves exactly as expected, gradually contracting as it moves towards core collapse.
In contrast, the BHs in Run 2 induce dramatic core expansion after
$\approx 650$ Myr. The plotted LMC/SMC data consists of all clusters from the WFPC2 
study of \citet{mackey:03a,mackey:03b} as well as preliminary ACS results
from Mackey et al. (2007b, in prep.) (program \#9891).
\vspace{-5mm}}
\label{f:pair1}
\end{figure}

The evolution of our non-MSeg runs is illustrated in Fig. \ref{f:pair1}. Run 1 constitutes 
the simplest case -- no primordial MSeg and no retained BHs. It behaves exactly as expected 
for a classical globular cluster. There is an early mass-loss phase ($\tau \la 100$ Myr) 
due to the evolution of the most massive cluster stars. During this phase, BHs are 
formed in supernova explosions between $3.5$-$10$ Myr; however, all receive large
velocity kicks and escape. The severe early mass-loss is not reflected in the evolution of 
$r_c$, presumably because it is evenly distributed throughout the cluster. Subsequently, the 
core radius slowly contracts as two-body relaxation proceeds and mass segregation 
sets in. The median relaxation time at $\tau = 10^8$ yr is $t_{rh} \sim 2$ 
Gyr. At $\tau_{{\rm max}} = 12$ Gyr $\approx 6\,t_{rh}$ the cluster has not yet entered 
the core-collapse phase.

Now consider Run 2, which is identical to Run 1 except that $f_{{\rm BH}} = 1$. 
Once early stellar evolution is complete, the BHs are more massive than all 
other cluster members (of mean mass $m_{*} \approx 0.5\,{\rm M}_\odot$) and are hence subject
to mass stratification on a time-scale of $\sim (m_{*} / m_{{\rm BH}})\, t_{rh} \approx 100$ Myr. 
By $200$ Myr, the mass density of the BHs within a radius of $0.5$ pc is already roughly 
equal to that of the stars; by $400$ Myr it is about three times larger. Soon after, the central 
BH subsystem becomes unstable to further contraction \citep[][ Eq. 3-55]{spitzer:87} and 
decouples from the stellar core in a runaway gravothermal collapse.
At $490$ Myr, the central density of 
the BH subsystem is $\sim 80$ times that of the stars. This is sufficient for the creation 
of stable BH binaries in three-body interactions -- the first is formed at $\sim 510$ Myr, and 
by $800$ Myr there are four. Until this phase, the evolution of Run 2 is observationally 
identical to that of Run 1. Neither BH retention, nor the subsequent formation of a central BH 
subsystem leads to differential evolution of $r_c$. This contrasts with the results of 
\citet{merritt:04} who found significant early expansion in their models due to the sinking
of BHs. We attribute this difference to the much higher degree of central mass concentration in 
their initially cusped clusters, which thereby respond more strongly and more rapidly to the 
perturbations induced by sinking remnants. These authors also noted the possibility of
further cluster expansion due to subsequent evolution of the BH subsystem. We indeed observe 
expansion due to such processes (see below).

Once formed, binary BHs undergo superelastic collisions with other BHs in the core.
The binaries become ``harder'', and the released binding energy 
is carried off by the interacting BHs. This leads to BHs being {\it scattered} outside 
$r_c$, often into the cluster halo, as well as to BHs being {\it ejected} from the cluster
(we retain this terminology henceforth).
Eventually a BH binary is sufficiently hard that the recoil velocity imparted to it
during a collision is larger than the cluster escape velocity, and the binary is ejected.
A BH scattered outside the cluster core gradually sinks back into the centre
via dynamical friction, thus transferring its newly-gained energy to the stellar component 
of the cluster. Most is deposited within $r_c$, where the stellar density is greatest. 
The ejection of BHs also transfers energy to the cluster, since a mass $m$ escaping 
from a cluster potential well of depth $|\Phi|$ does work $m|\Phi|$ on the cluster. 
This mechanism is particularly effective in heating the stellar core, since BHs 
are ejected from the very centre of the cluster, and the energy contributed to 
each part of the cluster is proportional to the contribution which that part makes to 
the central potential. 

Together, these two processes (scattering and ejection) result in significant core expansion, 
starting between $\tau \approx 600-700$ Myr. Expansion continues for the remainder of the 
simulation, which terminates at $\tau_{{\rm max}} \approx 10.6$ Gyr. The size of 
$r_c$ is roughly proportional to $\log \tau$, consistent with the upper envelope 
of the observed cluster distribution. However, in this model the expansion begins 
too late for the evolution to trace the upper envelope exactly; rather, it runs parallel.

The number of stable BH binaries in the system peaks at $5$ at $\tau \approx 890$ 
Myr. After this point, there are $0-5$ BH binaries at any given time. Single and binary
BHs are continually ejected; however, empirically, both escape rates depend logarithmically 
on $\tau$ -- i.e., $dN_{{\rm e}}/d\tau \propto \log\tau$.  
This arises due to the decreasing density of the central BH subsystem -- the number 
of BHs is falling because of ejections; these ejections also heat the BH core. The BH-BH 
encounter rate therefore decreases with time. Hence, the BH binary hardening rate decreases, 
as do the BH ejection rates. Furthermore, the stellar core is also less efficiently heated 
with time -- this is reflected in the roughly logarithmic dependence of $r_c$ on $\tau$. By 
$\tau_{{\rm max}} \approx 10.6$ Gyr, $96$ single BHs, $15$ binary BHs and one triple BH have
escaped; $65$ single BHs and $2$ binary BHs remain in the cluster. This is at odds 
with early studies \citep*[e.g.,][]{kulkarni:93,sigurdsson:93} which predicted depletion of 
BH populations on timescales much less than cluster lifetimes. The decreasing BH encounter rate 
seen in our models prolongs the life of the BH subsystem for much longer than previously appreciated.

The mean mass of stellar escapers is identical in both Runs 1 and 2, at $0.339\,{\rm M}_\odot$.
This is less than $m_{*}$ at all times. The distributions of velocities with which stars 
escape are also indistinguishable. These results imply that both models lose stars 
solely due to relaxation processes. There is only a tiny group of $\sim 30$ high velocity escapers 
in Run 2, indicating that stars interact closely with BH binaries only very rarely. Heating 
of the stellar component via close interactions is negligible -- the 
hardening of BH binaries is driven solely through interactions with other BHs.
At $\tau = 10$ Gyr, the masses of Runs 1 and 2, respectively, are $0.36\,M_{{\rm tot}}$ 
and $0.29\,M_{{\rm tot}}$, reflecting the fact that
Run 2 is more loosely bound than Run 1 for the majority of its evolution.

\vspace{-4mm}
\subsection{$N$-body pair 2: Strong mass segregation}
\label{ss:pair2}
\begin{figure}
\centering
\vspace{-2mm}
\includegraphics[width=8cm]{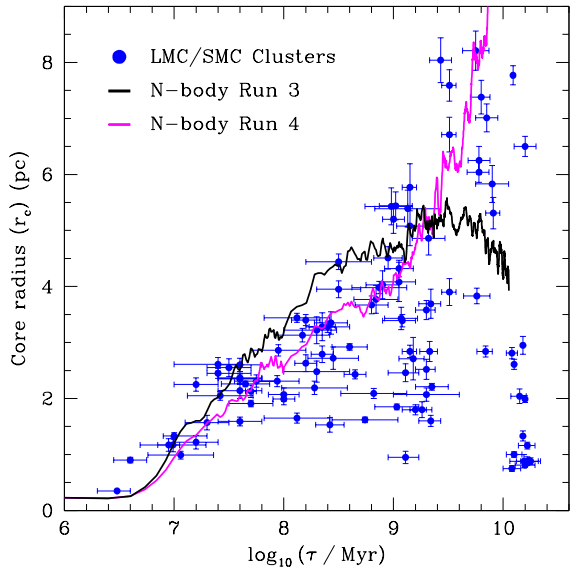}
\caption{Core radius evolution of Runs 3 and 4, which are initially identical,
with significant primordial MSeg. They have $f_{{\rm BH}} = 0$ and $1$, respectively.
Both expand dramatically at early times due to mass-loss from stellar evolution.
Subsequently, Run 3 begins to contract as two-body relaxation proceeds. In contrast, 
Run 4 continues expanding due to its BH population.\vspace{-5mm}}
\label{f:pair2}
\end{figure}

Runs 3 and 4 are primordially MSeg versions of Runs 1 and 2, respectively.
Early mass-loss due to stellar evolution is highly centrally concentrated --
hence the amount of heating per unit mass lost is maximised, leading to dramatic early 
core expansion (Fig. \ref{f:pair2}). Run 3 traces the observed upper envelope of clusters 
until several hundred Myr. Run 4 retains its BHs and hence loses less mass than Run 3 -- 
this is reflected in its smaller $r_c$. After the early mass-loss phase is complete, 
core expansion stalls in both runs. Two-body relaxation gradually takes over in Run 3, 
leading to a slow contraction in $r_c$. At $\tau = 1$ Gyr, $t_{rh} \approx 4$ Gyr; hence 
this cluster is not near core collapse by $\tau_{{\rm max}} \approx 11.2$ Gyr. At 
$\tau = 10$ Gyr, the remaining mass in Run 3 is $0.30\,M_{{\rm tot}}$.

In Run 4, the BH population evolves similarly to that in Run 2. One might naively expect 
the earlier development of a compact BH subsystem in Run 4, because the BHs are already 
located in the core due to the primordial MSeg. However, the centrally concentrated mass-loss 
hampers the accumulation of a dense BH core, and the first binary BH does not form until 
$570$ Myr, a similar time to the non-MSeg model. The BH subsystem 
evolves more slowly than that in Run 2 -- by $\tau_{{\rm max}} = 10$ Gyr, 
there are still $95$ single BHs and $2$ binary BHs remaining in the cluster. 
As in Run 2, the evolution of the BH subsystem leads to expansion of $r_c$. This begins
at $\tau \approx 800$ Myr and continues until $\tau_{{\rm max}}$. As previously, $r_c$
behaves roughly as $\log \tau$ during this phase. By $\tau_{{\rm max}}$, Run 4 has 
$r_c \sim 11$ pc, comparable to that observed for the most extended old Magellanic Cloud
clusters (e.g., Reticulum). However, it is only weakly bound, retaining 
$\sim 0.13\,M_{{\rm tot}}$. This mass loss is not the driver for the core expansion,
so a more massive cluster could have comparable expansion while retaining more of its 
total mass. Indeed, extended old LMC clusters typically have masses
$\approx 10^5\,{\rm M}_\odot$ \citep{mackey:03a}, which may easily be $\sim 0.1\,M_{{\rm tot}}$. 

\vspace{-5mm}
\section{Discussion}
\label{s:discussion}
Our four simulations cover the observed cluster distribution in radius-age space, 
thereby defining a dynamical origin for the radius-age trend. At ages less than a 
few hundred Myr, cluster cores expand due to centrally concentrated mass-loss from 
stellar evolution. At later times, expansion is induced via heating due to a BH 
population. Although early mass-loss may result in significant core expansion, a 
cluster cannot reach the upper right corner of the radius-age space by this means 
alone -- the mass-loss phase is too short, and the maximum allowed expansion rate 
during this phase is defined by the observed upper envelope of clusters. Only with
prolonged expansion due to BHs can $\sim 10$ Gyr old clusters with $r_c > 6$ pc 
be explained in this model. Although we have assumed $f_{{\rm BH}} = 1$,
full retention is not necessary for cluster expansion. BH kicks of order
$10 \la v_{{\rm kick}} \la 20$ km$\,$s$^{-1}$ would result in $f_{{\rm BH}} \sim 0.5$
in our models; we expect $r_c$ evolution in such systems to be intermediate
between that of Runs 1 and 2, or Runs 3 and 4. We will address this issue further 
in an upcoming paper (Mackey et al. 2007a, in prep.).

Galactic globular clusters, with $N \sim 10^6$, are an order of magnitude more massive 
than our present models. However, we expect the evolution described above to scale to 
such objects -- reflected in the fact they possess an $r_c$ distribution consistent with 
that observed for the oldest Magellanic Cloud clusters \citep{mackey:04}. This is because
the mass fraction of BHs formed in a cluster is dependent only on the IMF and minimum 
progenitor mass, neither of which should change with $M_{{\rm tot}}$, while a larger 
$M_{{\rm tot}}$ implies a larger $f_{{\rm BH}}$ 
since it is easier to retain newly-formed BHs. The densities 
in our models are consistent with those observed for globular clusters; hence we expect the 
same processes to operate on similar time-scales, although BHs are likely to be more
difficult to eject in more massive clusters -- increasing the potential of each BH to 
heat the cluster via additional scattering-sinking cycles. 
Core expansion due to mass-loss or BH heating has strong implications for the observed 
properties of Galactic globular clusters (e.g., the fraction which are core-collapsed) as 
well as their survivability. Extended clusters 
are significantly more susceptible to tidal disruption, so it is 
important to account for expansion effects in studies of the evolution of the globular 
cluster mass function, for example. Core expansion due to BHs may also offer a 
viable explanation for the origin of the luminous, unusually extended globular clusters 
found in M31, which are $>10$ Gyr old metal-poor objects \citep{mackey:06}.

Our model requires variations in BH population size between otherwise similar 
clusters. There are a variety of possibilities in this regard. First, the number 
of BH-forming stars in a cluster is small, so there will be sampling-noise variations 
between clusters. Further, any dispersion in stellar rotation may introduce mass-loss 
variations and further dispersion in BH numbers. Natal BH kicks are poorly constrained
at present -- typical estimates lie in the range $0 \la v_{{\rm kick}} \la 200$ km$\,$s$^{-1}$,
with kicks of a few tens of km$\,$s$^{-1}$ possibly favoured 
\citep[e.g.,][ and references therein]{willems:05}. Stellar binarity may therefore
play a significant role in retaining cluster BHs, as will the
initial cluster mass and degree of primordial MSeg, especially if 
$v_{{\rm kick}} \approx v_{{\rm esc}}$. Metallicity may also be a key factor,
as theory suggests that BH production is more frequent, and $m_{{\rm BH}}$ is greater for 
metal poor stars than for metal rich stars \citep[e.g.,][]{zhang:07}. In this respect,
the age-metallicity relationships of the Magellanic Clouds (where $[$Fe$/$H$]$
decreases for clusters of increasing age) may play a central role in shaping the
radius-age trend. Similarly, the spread in $[$Fe$/$H$]$ for Galactic globulars
may have been important in determining the structural properties of these objects.
Our results imply that clusters possessing significant BH populations are, for most
of their lives, low-density objects in which the timescale for close encounters 
between stars and BHs is very long. It is therefore unsurprising that no BH 
X-ray binaries are seen in the $\sim 150$ Galactic globulars \citep{verbunt:06}.

\vspace{-5mm}
\section*{Acknowledgments}
We thank Sverre Aarseth for the use of {\sc nbody4} and for his valuable assistance, 
Jarrod Hurley for his code to calculate the magnitudes of {\sc nbody4} stars, and Pete 
Bunclark and Mick Bridgeland for technical support with the IoA GRAPE-6. 
ADM is supported by a Marie Curie Excellence
Grant from the European Commission under contract MCEXT-CT-2005-025869. MIW 
acknowledges support from a Royal Society University Research Fellowship. MBD is
a Royal Swedish Academy Research Fellow supported by a grant from the Knut and
Alice Wallenberg Foundation.

\bsp

\label{lastpage}


\vspace{-5mm}
\begin{thebibliography}{99}
\bibitem[\protect\citeauthoryear{Aarseth}{2003}]{aarseth:03} 
  Aarseth S.J., 2003, Gravitational $N$-body Simulations. Cambridge University 
  Press, Cambridge
\bibitem[\protect\citeauthoryear{Bergeron, Wesemael \& Beauchamp}{Bergeron et al.}{1995}]{bergeron:95}
  Bergeron P., Wesemael F., Beauchamp A., 1995, PASP, 107, 1047
\bibitem[\protect\citeauthoryear{Bonnell \& Bate}{2006}]{bonnell:06}
  Bonnell I.A., Bate M.R., 2006, MNRAS, 370, 488
\bibitem[\protect\citeauthoryear{Casares}{2006}]{casares:06}
  Casares J., 2006, in Proc. IAU Symp. 238, Black Holes: from
  stars to galaxies, in press (astro-ph/0612312)
\bibitem[\protect\citeauthoryear{de Grijs et al.}{2002}]{degrijs:02}
  de Grijs R., Gilmore G.F., Johnson R.A., Mackey A.D., 2002, MNRAS, 331, 245
\bibitem[\protect\citeauthoryear{Elson, Fall \& Freeman}{Elson et al.}{1987}]{elson:87}
  Elson R., Fall S.M., Freeman K.C., 1987, ApJ, 323, 54
\bibitem[\protect\citeauthoryear{Elson, Freeman \& Lauer}{Elson et al.}{1989}]{elson:89}
  Elson R., Freeman K.C., Lauer T.R., 1989, ApJ, 347, L69
\bibitem[\protect\citeauthoryear{Hurley, Pols \& Tout}{Hurley et al.}{2000}]{hurley:00}
  Hurley J.R., Pols O.R., Tout C.A., 2000, MNRAS, 315 543
\bibitem[\protect\citeauthoryear{Hurley, Tout \& Pols}{Hurley et al.}{2002}]{hurley:02}
  Hurley J.R., Tout C.A., Pols O.R., 2002, MNRAS, 329, 897
\bibitem[\protect\citeauthoryear{Hunter et al.}{2003}]{hunter:03}
  Hunter D.A., Elmegreen B.G., Dupuy T.J., Mortonson M., 2003, AJ, 126, 1836
\bibitem[\protect\citeauthoryear{Kroupa}{2001}]{kroupa:01}
  Kroupa P., 2001, MNRAS, 322, 231
\bibitem[\protect\citeauthoryear{Kulkarni, Hut \& McMillan}{Kulkarni et al.}{1993}]{kulkarni:93}
  Kulkarni S.R., Hut P., McMillan S., 1993, Nature, 364, 421
\bibitem[\protect\citeauthoryear{Kurucz}{1992}]{kurucz:92}
  Kurucz R.L., 1992, in Barbuy B., Renzini A., eds., Proc. IAU Symp. 149, 
  The Stellar Populations of Galaxies. Kluwer, Dordrecht, p. 225
\bibitem[\protect\citeauthoryear{Mackey \& Gilmore}{2003a}]{mackey:03a} 
  Mackey A.D., Gilmore G.F., 2003a, MNRAS, 338, 85
\bibitem[\protect\citeauthoryear{Mackey \& Gilmore}{2003b}]{mackey:03b} 
  Mackey A.D., Gilmore G.F., 2003b, MNRAS, 338, 120
\bibitem[\protect\citeauthoryear{Mackey \& Gilmore}{2004}]{mackey:04} 
  Mackey A.D., Gilmore G.F., 2004, MNRAS, 355, 504
\bibitem[\protect\citeauthoryear{Mackey et al.}{2006}]{mackey:06}
  Mackey A.D., et al., 2006, ApJ, 653, L105
\bibitem[\protect\citeauthoryear{Makino et al.}{2003}]{makino:03} 
  Makino J., Fukushige T., Koga M., Namura K., 2003, PASJ, 55 ,1163
\bibitem[\protect\citeauthoryear{McLaughlin \& van der Marel}{2005}]{mclaughlin:05}
  McLaughlin D., van der Marel R., 2005, ApJS, 161, 304
\bibitem[\protect\citeauthoryear{Merritt et al.}{2004}]{merritt:04}
  Merritt D., Piatek S., Portegies Zwart S., Hemsendorf M., 2004, ApJ, 608, L25
\bibitem[\protect\citeauthoryear{Meylan \& Heggie}{1997}]{meylan:97}
  Meylan G., Heggie D.C., 1997, A\&AR, 8, 1
\bibitem[\protect\citeauthoryear{Sigurdsson \& Hernquist}{1993}]{sigurdsson:93}
  Sigurdsson S., Hernquist L., 1993, Nature, 364, 423
\bibitem[\protect\citeauthoryear{Spitzer}{1987}]{spitzer:87}
  Spitzer L., 1987, Dynamical Evolution of Globular Clusters. Princeton University
  Press, Princeton
\bibitem[\protect\citeauthoryear{Verbunt \& Lewin}{2006}]{verbunt:06}
  Verbunt F., Lewin W., 2006, in Lewin W., van der Klis M., eds., Compact stellar X-ray sources.
  Cambridge University Press, Cambridge, p. 341
\bibitem[\protect\citeauthoryear{Wilkinson et al.}{2003}]{wilkinson:03}
  Wilkinson M.I., Hurley J.R., Mackey A.D., Gilmore G.F., Tout C.A., 2003, MNRAS, 343, 1025
\bibitem[\protect\citeauthoryear{Willems et al.}{2005}]{willems:05}
  Willems B., et al., 2005, ApJ, 625, 324
\bibitem[\protect\citeauthoryear{Zhang, Woosley \& Heger}{Zhang et al.}{2007}]{zhang:07}
  Zhang W, Woosley S., Heger A., 2007, ApJ, submitted (astro-ph/0701083)
\end{thebibliography}
\end{document}